\documentclass[final,5p,times,twocolumn]{elsarticle}

\usepackage{amssymb}
\usepackage{amsmath}
\usepackage{gensymb}

\begin{document}

\begin{frontmatter}



\title{A review of the classical Doppler effect based on the mathematical description of the phase function}


\author[first]{\'{O}scar Alejos and Jos\'{e} Mar\'{\i}a Mu\~{n}oz}
\affiliation[first]{organization={University of Valladolid},
            addressline={Department of Electricity and Electronics}, 
            city={Valladolid},
            postcode={47011}, 
            country={Spain}}

\begin{abstract}
The classical Doppler effect is explicitly reformulated by determining the phase function associated with a moving source. General expressions for frequency shift and aberration are derived and compared with the results obtained from a geometrical treatment, showing complete consistency between both approaches. The phase formalism provides direct expressions for the local frequency and the wave vector through the temporal and spatial derivatives of the phase, respectively. In this framework, aberration appears as a consequence of the local propagation direction, while the Doppler shift is determined by the temporal evolution of the phase perceived by the observer. The resulting wave vector naturally incorporates the effects of source motion and leads to a compact formulation of the general Doppler effect in terms of relative velocities. This approach yields a unified description of moving sources, moving observers, and moving propagation media, and can be applied in arbitrary reference frames.
\end{abstract}




\end{frontmatter}




\section{Introduction}
\label{introduction}

The Doppler effect (DE) has been extensively studied after its discovery by Christian Doppler in 1842. Since the original work of Doppler \cite{Doppler:03}, the phenomenon has become one of the classical subjects of wave physics and acoustics, receiving extensive treatment in foundational works such as Rayleigh's {\em The Theory of Sound} \cite{rayleigh:96}. Many textbooks \cite{cutnell:13,halliday:13,serway:18,tipler:22} include discussions of this effect, both in its classical and relativistic versions.
In the classical acoustic Doppler effect, the existence of a material propagation medium introduces features that are absent in the relativistic treatment of electromagnetic waves in vacuum. Consequently, the velocities of the source, the observer, and the medium must be distinguished explicitly, as discussed in standard acoustics references such as Pierce's {\em Acoustics: An Introduction to Its Physical Principles and Applications} \cite{pierce:19} or Blackstock's {\em Fundamentals of Physical Acoustics} \cite{blackstock:00}. Nevertheless, most textbook presentations are generally restricted to particular geometries of motion and do not explicitly formulate the problem in terms of the phase function of the propagating wave. In fact, only the situation in which the observer is within the trajectory of the emitter is considered in most of them, that is, the positions of the emitter and the observer are collinear with the emitter's trajectory. The non-collinear case is indeed the subject of recent scientific papers \cite{Bokor:09,Klinaku:19,Klinaku:21,Michel:22}, where even some attempts to unify the classical and relativistic DEs are made. These approaches are generally based on geometrical constructions. However, some presented results do not adequately account for the nonlinear dependence of phase on time at the observer's location. The reason is that they compare one period of the emitted signal with what is called a period perceived by the observer, that is, the interval of time between two identical vibration states of the wave, a question that will be addressed along this paper through the explicit determination of the phase function. Moreover, some treatments available in the acoustics literature, while perfectly adequate for many practical applications, do not explicitly address the time-dependent phase evolution perceived by an observer in the general non-collinear configuration \cite{pierce:19}.

A more general wave-theoretical description is obtained by introducing the phase function of the propagating wave, where the propagation of a wave field can be described through a phase function whose temporal and spatial derivatives define the local frequency and wave vector, respectively. This viewpoint is widely used in geometrical optics and wave theory, particularly in the context of slowly varying wave fields and eikonal descriptions \cite{whitham:74}. The present work applies this framework to the classical DE generated by moving sources and observers, showing that frequency shift and aberration emerge naturally from the phase function of the propagating wave. Accordingly, the present paper attempts to give a clearer view of classical DE, in particular by first contrasting the results provided by a careful revision of the geometrical description with those resulting from an explicit phase-based formulation in terms of the spatial and temporal dependence of the phase of the propagating wave in the medium. The phase function is then used to address the more general case in which the emitter, observer, and medium are in motion in the reference frame under consideration, a situation for which a purely geometrical treatment becomes increasingly cumbersome.

The work is organized according to the following structure. First, the case of a stationary observer in the reference frame in which the propagating medium is at rest is reviewed, emitter moving at constant velocity in that reference frame. Initially, the problem is approached by a purely geometrical construct, which justifies the aberration effect when the observer is not in the path of the emitter. The corresponding expression for the frequency shift, defined as the instantaneous variation of the phase perceived by the observer, is also obtained. In the same section, the problem is addressed in a much more intuitive way, that is, by constructing a function that defines the phase at each instant and at each point in space in the observer's reference frame. Aberration and frequency shift appear naturally from the derivatives of the phase function with respect to time and position, verifying the consistency between this approach and the previous one. In the following section, the case of an observer in motion in a reference frame in which the medium of propagation and the emitter are at rest is briefly analyzed. Finally, the more general case of moving observer and emitter is addressed. The study presents the expression for the frequency shift due to DE in terms of relative positions and velocities, which can be easily extended to any reference frame, particularly those in which the propagation medium is not at rest and the wavefronts undergo a drift due to the medium velocity.  

\section{Moving emitter}
\subsection{Geometrical construct}
\begin{figure}[htbp]
\centering
\includegraphics[scale=0.5]{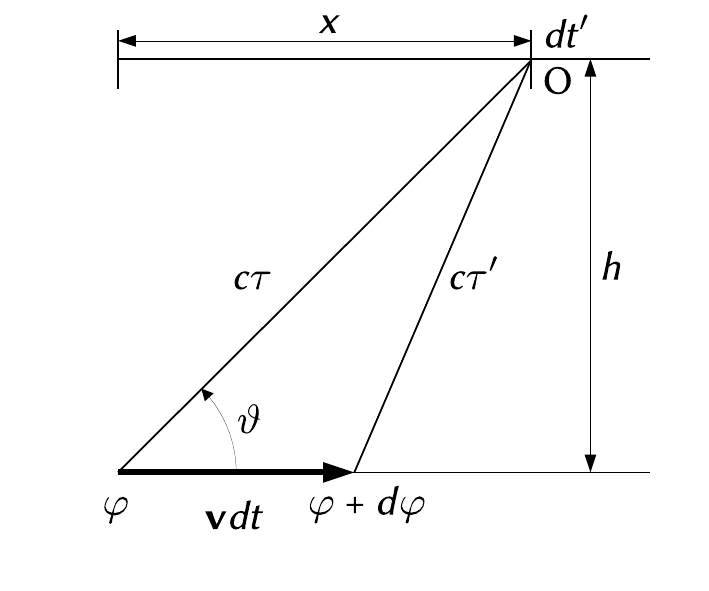}
\caption{Geometry of the problem of a moving emitter. To facilitate the viewing of the drawing, the distance $vdt$ has been exaggeratedly enlarged.}\label{Fig:01}
\end{figure}

Consider that, at a given time instant $t$, the emitter is oscillating and its phase is $\varphi$. Without loss of generality, the observer can be located at the origin of coordinates in its reference frame so that the emitter at that instant is at the position given by a certain $x$-coordinate and the distance $h$ between the observer and the line defining the trajectory of the emitter, as shown in fig. \ref{Fig:01}. Assume that at that instant the distance between the emitter and the observer is $c\tau$, where $c$ is the propagation velocity in the medium, which will be taken to be at rest in the observer's reference frame. In fact, $\left(c\tau\right)^2=x^2+h^2$.

The emitter moves with velocity $v$ in the observer's reference frame, so that it travels a distance $vdt$ after a time $dt$. At instant $t+dt$, the phase of the emitter will be $\varphi+d\varphi$, with $d\varphi=2\pi fdt$, $f$ being the frequency of the emitter. This new phase will be measured at the observer's position one time interval $\tau'$ later. The phase measured at the position of the observer changes in a quantity $d\varphi$ after a time $dt'$, which can be computed as $dt'=\tau'+dt-\tau$. Since, at instant $t+dt$, the emitter is located at the position $x+vdt$, then $\left(c\tau'\right)^2=(x+vdt)^2+h^2$. According to this,
\begin{equation}
\begin{aligned}
dt'&=\tau'+dt-\tau=\\ 
&=\frac{\sqrt{\left(x+vdt\right)^2+h^2}}{c}+dt-\frac{\sqrt{x^2+h^2}}{c}\approx\\
&\approx dt\left(1+\frac{x\frac{v}{c}}{\sqrt{x^2+h^2}}\right)\text{.}
\end{aligned}
\end{equation}
The instant frequency $f'$ measured at the observer's position is then given by $f'=\frac{1}{2\pi}\frac{d\varphi}{dt'}$. After renaming $\beta=\frac{v}{c}$ and rewriting $\cos\vartheta=-\frac{x}{\sqrt{x^2+h^2}}$, the result
\begin{equation}
f'=\frac{f}{1-\beta\cos\vartheta}\text{.}\label{eqn:02}
\end{equation}
is obtained. As a consequence of this expression, all observers lying along a direction characterized by the angle $\vartheta$ perceive the same frequency $f'$, but at different time instants, depending on the distance from each observer to the emitter along this line. The previous expression coincides with the Galilean limit of the relativistic Doppler formula. It is sufficient to consider that in this expression all times are referred to observer's frame, which is the same as that of the reference frame of the emitter in the framework of Galilean relativity. The missing factor $\frac{1}{\gamma}=\sqrt{1-\beta^2}$ in the expression (\ref{eqn:02}) appears naturally if one considers the time dilation associated with the change in reference frame within special relativity.

Aberration arises because the emitter changes its position during the propagation time of the wavefront, that is, the sound comes from one direction but the emitter is seen in a different direction. Both directions are related to each other.

\begin{figure}[htbp]
\centering
\includegraphics[scale=0.5]{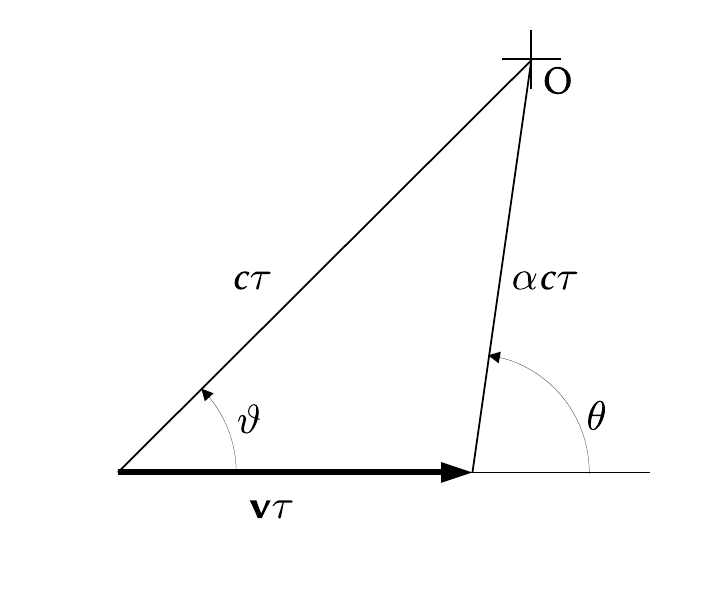}
\caption{Aberration. The direction from which the sound comes and the direction in which the emitter is seen are different.}\label{Fig:02}
\end{figure}

Consider now fig. \ref{Fig:02}. When the observer perceives the frequency $f'$ coming from the direction given by the angle $\vartheta$, a time $\tau$ has elapsed since the phase variation of the emitter produced such a sound, due to the distance $c\tau$ at which the emitter was located from the observer. In that time $\tau$, the emitter has moved a distance $v\tau$, and the distance the emitter is from the observer can be written as a certain factor $\alpha$ of the distance $c\tau$. Applying the cosine law, it is found that
\begin{equation}
\alpha^2+\beta^2+2\alpha\beta\cos\theta=1\text{.}\label{eqn:03}
\end{equation}
Additionally, sine law establishes that
\begin{equation}
\sin\vartheta=\alpha\sin\theta\text{.}\label{eqn:04}
\end{equation}

Both equations can be combined to give the desired result linking both angles $\vartheta$ and $\theta$, as demonstrated in \ref{app:A}. In this way,
\begin{subequations}
\begin{align}
\sin\vartheta &=\sin\theta\left(\sqrt{1-\beta^2\sin^2\theta}-\beta\cos\theta\right)\text{,}\label{eqn:05a}\\
\cos\vartheta &=\beta\sin^2\theta+\cos\theta\sqrt{1-\beta^2\sin^2\theta}\text{.}\label{eqn:05b}
\end{align}
\end{subequations}

Now, expression (\ref{eqn:02}) can be written in its final form using (\ref{eqn:05b}), showing the frequency shift due to DE, that is,
\begin{equation}
f'=\frac{f}{1-\beta\left(\beta\sin^2\theta+\cos\theta\sqrt{1-\beta^2\sin^2\theta}\right)}\text{.}\label{eqn:06}
\end{equation}
The expression above reduces to the widely known formulae for collinear displacement of the emitter with the line joining the emitter and the observer, that is, for $\theta$ equal to either $0$ (emitter approaching the observer): 
\begin{equation}
f'=\frac{f}{1-\beta}\text{,}
\end{equation}
or $180\degree$ (emitter moving away from the observer):
\begin{equation}
f'=\frac{f}{1+\beta}\text{.}
\end{equation}
Additionally, DE vanishes when $\vartheta = 90\degree$, and the observer sees the emitter at a certain angle $\theta_v\neq 90\degree$ given by the condition 
\begin{equation}
\sin\theta_v=\frac{1}{\sqrt{1+\beta^2}}\text{,}\label{eqn:09}
\end{equation}
once the observer has been overtaken by the emitter.

\subsection{Instantaneous phase determination}

The phase-based approach adopted here is rooted in the general idea that the wave field may be completely characterized by a scalar phase function. In this context, the construction of the wave field generated by a moving source follows the standard treatment of acoustic radiation and retarded propagation times discussed in classical acoustics \cite{morse:68}, while local propagation properties can be obtained directly from the derivatives of the phase function, an approach widely employed in wave theory and geometric optical approximations \cite{whitham:74}.

Since in the framework of Galilean relativity, the wave equation is not invariant with respect to coordinate transformations, the calculation of the phase of a wave in a reference frame in which the emitter is in motion with respect to the observer cannot be performed by simply considering circular wave fronts in emitter's reference frame and applying the corresponding coordinate transformation. The procedure is therefore not straightforward but must be approached from the wavefronts created by the emitter at each of the points it travels through.

To determine the phase of each point in space, it is then necessary to first estimate the position $\mathbf{x}_0$ where the emitter was located when the corresponding wavefront that reached that point was emitted. The time $\tau$ the wavefront requires to travel the distance from $\mathbf{x}_0$ to the point is given by the quotient between that distance and the phase velocity. {\em The phase at the point at any time $t$ is indeed equal to the phase the emitter had when it was at position $\mathbf{x}_0$, a time $\tau$ ago.} If $\varphi_e\left(t\right)$ is the current phase of the emitter, the phase at any point is now given by the function $\varphi_e\left(t-\tau\right)$, $\tau$ being a function of $t$ determined by the location of the point.

\begin{figure}[htbp]
\centering
\includegraphics[scale=0.5]{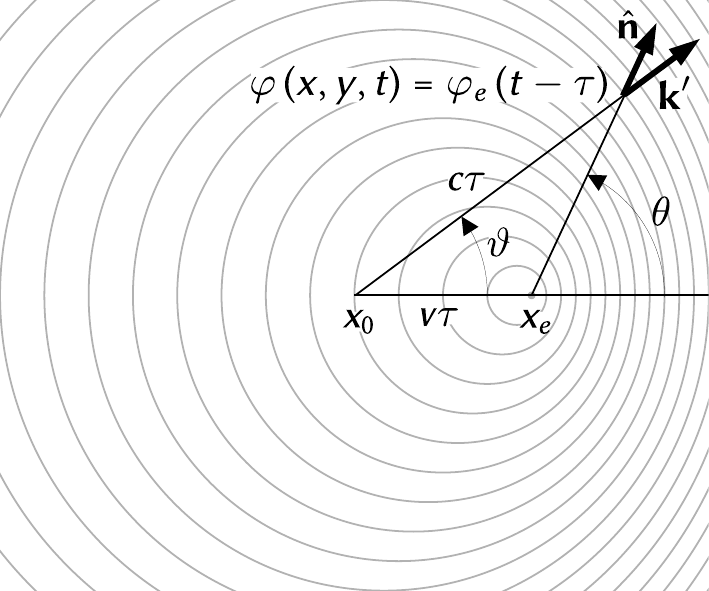}
\caption{A train of wavefronts. The phase at a point $\left(x,y\right)$ depends on the position $\mathbf{x}_0$ at which the corresponding wavefront was emitted, and the time elapsed until the front reaches that point. In that time, the emitter has advanced a certain distance until it reaches position $\mathbf{x}_e$. The figure shows the case of the constant velocity emitter.  
The angle $\theta$ defines the direction in which the observer sees the emitter at the instant $t$.}\label{Fig:03}
\end{figure}

According to the above discussion, the problem can be formulated through a pair of equations in the case of one emitter moving at constant velocity, as shown in fig. \ref{Fig:03}. The emitter can be considered to move along the x-axis. A certain wavefront reaches the point of coordinates $\left(x,y\right)$ at instant $t$, when the emitter is at a point of coordinates $\left(x_e,0\right)$. The wavefront was emitted when the emitter was at the position of coordinates $\left(x_0,0\right)$, a time $\tau$ ago. In this way,
\begin{subequations}
\begin{align}
x_e-x_0&=v\tau\text{,}\\
\left(x-x_0\right)^2+y^2&=\left(c\tau\right)^2\text{.}
\end{align}
\end{subequations}
By combining both equations, the following quadratic equation is obtained
\begin{equation}
\left(c^2-v^2\right)\tau^2-2\left(x-x_e\right)v\tau-\left[\left(x-x_e\right)^2+y^2\right]=0\text{,}
\end{equation}
that can be solved by selecting the physically meaningful positive root in such a way that $\tau$ equals  
\begin{equation}
\tau=\frac{1}{c}\frac{1}{1-\beta^2}\left[\beta\left(x-x_e\right)+\sqrt{\left(x-x_e\right)^2+\left(1-\beta^2\right)y^2}\right]\text{.}
\end{equation}

Without loss of generality, the instantaneous phase of the emitter can be written as $\varphi_e\left(t\right)=2\pi ft$. As previously stated, the phase $\varphi$ at any time $t$ at a point of coordinates $\left(x,y\right)$ is given by $\varphi_e\left(t-\tau\right)$, thus resulting in the expression of the dependence of the phase on position and time
\begin{equation}
\varphi\left(x,y,t\right)=2\pi ft-k\frac{\beta\left(x-vt\right)+\sqrt{\left(x-vt\right)^2+\left(1-\beta^2\right)y^2}}{1-\beta^2}\text{.}\label{eqn:13}
\end{equation}

\begin{figure*}[htbp]
\centering
\includegraphics[scale=1.2]{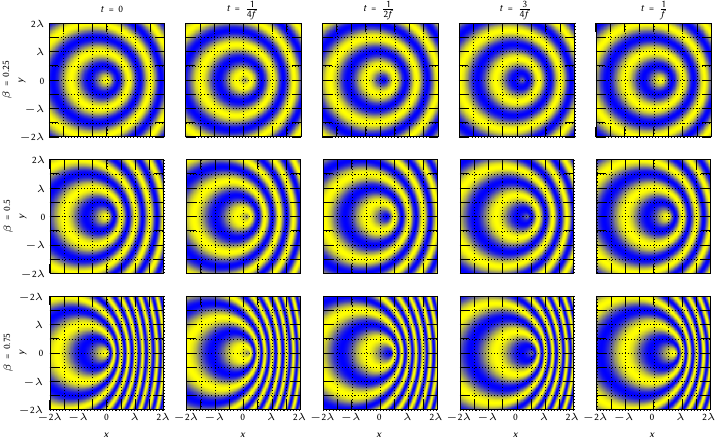}
\caption{Wavefronts generated by a moving emitter at different instants and for different velocities of the emitter (see supplementary videos 1 to 3). Check the sphericity of wavefronts.}
\label{Fig:04}
\end{figure*}

Figure \ref{Fig:04} displays the wavefronts generated by a moving emitter using color maps. The advance of the fronts is shown at different time instants within an interval equal to one oscillation period, given by $\frac{1}{f}$, and for different velocities of the emitter. The perfectly spherical shape of the wavefronts can be verified in all cases. 

Once the function that establishes the phase as a function of position and time has been determined, the propagation characteristics, that is, frequency and wave vector, can be easily derived. To do so, it suffices to consider $f'=\frac{1}{2\pi}\frac{d\varphi}{dt}$ for the former and $\mathbf{k}'=-\nabla\varphi$ for the latter, in agreement with the general phase formalism used in wave-propagation theory \cite{whitham:74}. Starting with the frequency, it turns out that
\begin{equation}
f'=f\left\{1+\frac{\beta}{1-\beta^2}\left[\beta+\frac{x-vt}{\sqrt{\left(x-vt\right)^2+\left(1-\beta^2\right)y^2}} \right]\right\}\text{.}
\end{equation}
which explicitly shows the time dependence of the {\em instantaneous} frequency perceived by the observer. This expression can be rewritten in terms of the angle $\theta$ defined in Figs. \ref{Fig:02} and \ref{Fig:03}, which yields
\begin{equation}
f'=f\left[1+\frac{\beta}{1-\beta^2}\left(\beta+\frac{\cos\theta}{\sqrt{1-\beta^2\sin^2\theta}} \right)\right]\text{.}\label{eqn:12}
\end{equation}
Proper handling of the above expression (see \ref{app:B}) leads to a formulation of the expression as follows
\begin{equation}
f'=\frac{f}{1-\beta\left(\beta\sin^2\theta+\cos\theta\sqrt{1-\beta^2\sin^2\theta}\right)}\text{,}
\end{equation}
that is, the result given in (\ref{eqn:06}).

In addition, the wave vector can be obtained as
\begin{equation}
\begin{aligned}
\mathbf{k}'&=\frac{k}{\left(1-\beta^2\right)\sqrt{1-\beta^2\sin^2\theta}}\times\\
&\times\left[\left(\beta\sqrt{1-\beta^2\sin^2\theta}+\cos\theta\right)\hat{\mathbf{x}}+\left(1-\beta^2\right)\sin\theta\hat{\mathbf{y}}\right]\text{,}\label{eqn:17}
\end{aligned}
\end{equation}
This expression can be straightforwardly rewritten in terms of the angle $\vartheta$ in fig.\ref{Fig:02} (see \ref{app:C}), which results in
\begin{equation}
\mathbf{k}'=\frac{k}{1-\beta\cos\vartheta}\left(\cos\vartheta\hat{\mathbf{x}}+\sin\vartheta\hat{\mathbf{y}}\right)\text{,}\label{eqn:18}
\end{equation}
as can be expected, since the direction of propagation at each point is perpendicular to the wavefronts, then pointing in the direction of the segment joining this point and that where the wavefront was started. Aberration results from the misalignment between the wave vector $\mathbf{k}'$ and the line joining emitter and observer represented by the vector $\hat{\mathbf{n}}$ depicted in fig.\ref{Fig:03}. Additionally, the observed and emitted wavelenghts, that is, $\lambda'$ and $\lambda$, fulfill the expected relationship
\begin{equation}
\lambda' f'=\lambda f=c
\end{equation}

\begin{figure}[htbp]
\centering
\includegraphics[scale=1]{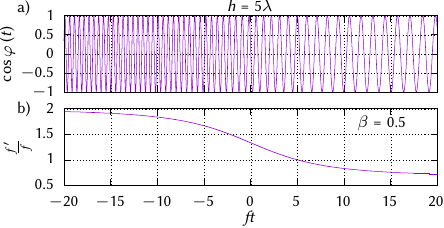}
\caption{DE for a moving emitter, passing by an observer located at a distance equal to $5\lambda$ to emitter's trajectory with a velocity given by $\beta=0.5$: a) temporal evolution of phase vibration at observer's location, and b) {\em instantaneous} frequency perceived by the observer. Time is expressed in nondimensional units.}\label{Fig:05}
\end{figure}

Figure \ref{Fig:05} shows a first example of these results. An emitter moving with velocity given by $\beta=0.5$ passes by an observer located at a distance of $5\lambda$ away from the emitter's trajectory. The time in these graphs is taken as non-dimensional by multiplying it by the emitter's frequency $f$, and the emitter and observer are at the closest positions when $t=0$. The dependence of the phase on time is shown in a), where shorter wavelenghts can be seen at the initial instants and these wavelengths become larger as the emitter progresses. Figure \ref{Fig:05}.b) shows how the {\em instantaneous} frequency perceived by the observer {\em decreases} in time. Aberration shifts the instant at which the observed frequency equals the emitted frequency since the observer perceives a frequency equal to the emitter's frequency not when the emitter and observer are at the closest positions $\left(\theta=90\degree\right)$, but at a later time instant given by $t=\frac{h}{c}=\frac{h}{\lambda f}$, that is, when $\vartheta=90\degree$.

\begin{figure}[htbp]
\centering
\includegraphics[scale=1]{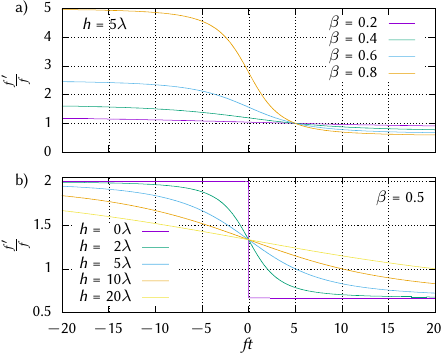}
\caption{DE for a moving emitter, passing by an observer. {\em Instantaneous} frequency perceived by an observer: a) located at a distance equal to $5\lambda$ to emitter's trajectory, when the emitter moves with different velocities, and b) located at different distance from emitter's trajectory, when the emitter moves with velocity $\beta=0.5$. (time expressed in nondimensional units).}\label{Fig:06}
\end{figure}

Figure \ref{Fig:06} shows more examples of these results. In case a) the observer is at a distance equal to $5\lambda$ away from the emitter's trajectory, who is moving at different velocities. Since the distance holds for all cases, the instant at which the frequency shift vanishes remains constant by the condition $ft=\frac{h}{\lambda}=5$, that is, the emitter is farther from the observer as the velocity increases, according to (\ref{eqn:09}).
In addition, frequency variation becomes sharper as the emitter's trajectory approaches the observer's position, as shown in b).

\section{Moving observer}
\subsection{Geometrical construct}
\begin{figure}[htbp]
\centering
\includegraphics[scale=0.5]{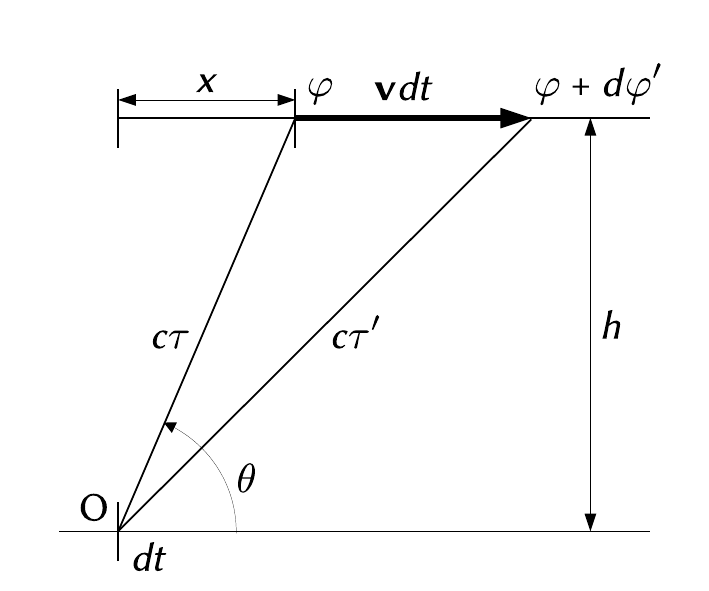}
\caption{Geometry of the problem of a moving observer. Again, the distance $vdt$ has been exaggeratedly enlarged to facilitate the viewing of the drawing.}\label{Fig:07}
\end{figure}

Consider now the case of a moving observer. The emitter is at rest in the reference frame where the propagating medium is still. With no loss of generality, the emitter has been located at the origin of the coordinates. At a given time instant $t$, the observer perceives a phase $\varphi$. Thus, the phase of the emitter is $\varphi+kc\tau$ at that instant, according to the drawing in fig.\ref{Fig:07}. After a time interval of $dt$, the observer has moved a distance $vdt$ and perceives a different phase $\varphi+d\varphi'$. The phase of the  emitter at that instant must be $\varphi+kc\tau+2\pi fdt$. Taking into account the new distance $c\tau'$ between the observer and the emitter, the equation $\varphi+d\varphi'=\varphi+kc\tau+2\pi fdt-kc\tau'$ holds, so that
\begin{equation}
\begin{aligned}
d\varphi'&=2\pi f dt-kc\left(\tau'-\tau\right)=\\ 
&=2\pi f \left(dt-\frac{\sqrt{\left(x+vdt\right)^2+h^2}}{c}+\frac{\sqrt{x^2+h^2}}{c}\right)\approx\\
&\approx 2\pi fdt\left(1-\frac{v}{c}\frac{x}{\sqrt{x^2+h^2}}\right)\text{.}
\end{aligned}
\end{equation}

The frequency in the observer's position is obtained as
\begin{equation}
f'=\frac{1}{2\pi}\frac{d\varphi'}{dt}=f\left(1-\beta'\cos\theta\right)\text{,}
\end{equation}
where $\beta'$ is defined considering the velocity $v=\beta' c$ of the observer. This is indeed the DE expected expression for moving observers. In this case, no aberration must be considered, since wave vectors are directed radially from the emitter, being then perpendicular to the wave fronts, as will be stated in the next subsection.

\subsection{Instantaneous phase approach}
\begin{figure}[htbp]
\centering
\includegraphics[scale=0.5]{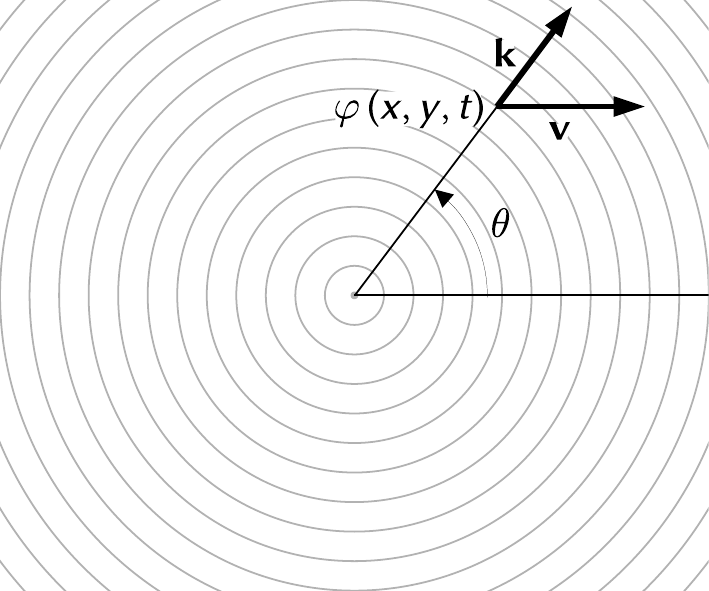}
\caption{DE for a moving observer, passing by an observer. The observer moves with velocity $\mathbf{v}$ and perceives the wave fronts moving in the direction of the wave vector $\mathbf{k}$.}\label{Fig:08}
\end{figure}

Within the emitter's reference frame, the medium is still, so that (\ref{eqn:13}) is rewritten as
\begin{equation}
\varphi\left(x,y,t\right)=2\pi ft-k\sqrt{x^2+y^2}\text{.}
\end{equation}
leading to concentric wave fronts centered at the emitter's location, as shown in fig.\ref{Fig:08}. The frequency $f'$ perceived by the observer is obtained from the derivative of the previous expression over time. The use of the convective derivative $\frac{d}{dt}=\frac{\partial}{\partial t}+\mathbf{v}\cdot\nabla$ is now necessary to yield again $f'=f\left(1-\beta'\cos\theta\right)$, as in the previous subsection. Taking into account that the application of the operator $\nabla$ results in the wave vector $\mathbf{k}=-\nabla\varphi=k\left(\cos\theta\hat{\mathbf{x}}+\sin\theta\hat{\mathbf{y}}\right)$, the frequency shift can be simply rewritten as
\begin{equation}
f'=f-\frac{1}{2\pi}\mathbf{k}\cdot\mathbf{v}\text{.}\label{eqn:23}
\end{equation}

\section{Moving emitter and observer}
\begin{figure}[htbp]
\centering
\includegraphics[scale=0.5]{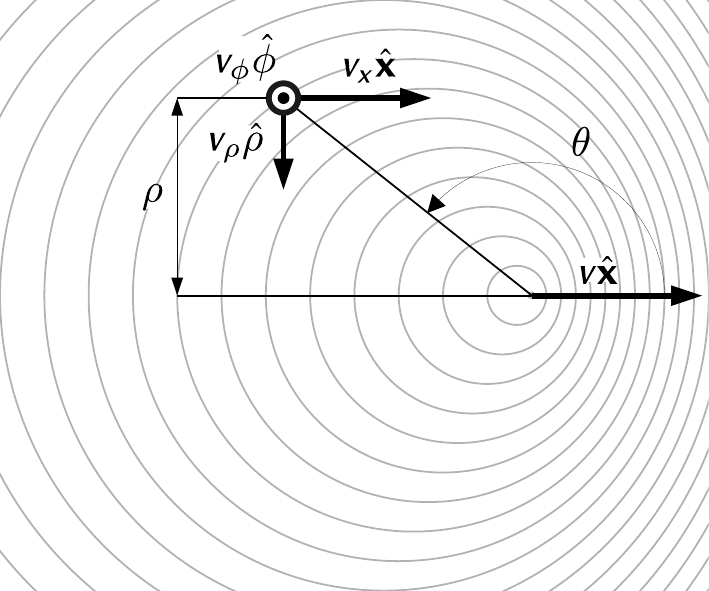}
\caption{DE for the case when both emitter and observer are moving with respect to the reference frame where the propagating medium is at rest. The velocity of the emitter is given by vector $v\hat{mathbf{x}}$, while the observer moves with velocity $v_x\hat{\mathbf{x}}+v_\rho\hat{\mathbf{\rho}}+v_\phi\hat{\mathbf{\phi}}$. The angle $\theta$ defining the relative instantaneous position between observer and emitter is also displayed.}\label{Fig:09}
\end{figure}

In contrast to electromagnetic waves in vacuum, the propagation of sound waves requires a material medium, whose state of motion plays a fundamental role in the classical description of DE \cite{pierce:19}. In this case, we will omit the geometric description due to its complexity and instead use the phase function directly. As in the previous sections, the velocities will be taken with respect to the reference frame in which the propagating medium is at rest, so that, without loss of generality, the emitter will be considered to move with velocity $v\hat{\mathbf{x}}$ with $v=\beta c$. However, now the motion of the observer must be written in the most general way as $v_x\hat{\mathbf{x}}+v_\rho\hat{\mathbf{\rho}}+v_\phi\hat{\mathbf{\phi}}$. The coordinate $\rho$ now plays the role of the coordinate $y$ in previous approaches, as shown in fig.\ref{Fig:09}. The use of the convective derivative yields the expression for the frequency shift as
\begin{equation}
f'=f\left[1-\frac{\beta\left(\beta_x-\beta\right)+\frac{\left(\beta_x-\beta\right)\cos\theta+\beta_\rho\left(1-\beta^2\right)\sin\theta}{\sqrt{1-\beta^2\sin^2\theta}}}{1-\beta^2}\right]\text{,}\label{eqn:24}
\end{equation}
where $\beta_x=\frac{v_x}{c}$ and $\beta_\rho=\frac{v_\rho}{c}$, and $\theta$ being the angle between the emitter's trajectory and the relative instantaneous position between the observer and the emitter depicted in fig.\ref{Fig:09}. Equation (\ref{eqn:24}) shows that the Doppler shift is independent of the azimuthal velocity component $v_\phi$, that is, as long as the observer goes round the line defining the emitter's trajectory keeping the same $\rho$ distance. This is consistent with the fact that the azimuthal motion does not modify the radial component of the relative velocity. Furthermore, the DE vanishes when both the observer and the emitter move parallel with the same velocity, when $\beta=\beta_x$ and $\beta_\rho=0$. Equation (\ref{eqn:24}) depends exclusively on relative positions, through the $\theta$-angle, and relative velocities, namely the velocity of the emitter with respect to the medium and the relative velocity between the observer and emitter. This result is then valid for any other reference frame. Hence, by defining $\left(\beta'_x,\beta'_\rho\right)$ as the components of the relative velocity of the observer with respect to the emitter, the previous equation can be rewritten as
\begin{equation}
f'=f\left[1-\frac{\beta\beta'_x\sqrt{1-\beta^2\sin^2\theta}+\beta'_x\cos\theta+\beta'_\rho\left(1-\beta^2\right)\sin\theta}{\left(1-\beta^2\right)\sqrt{1-\beta^2\sin^2\theta}}\right]\text{,}\label{eqn:25}
\end{equation}
where $\beta$ will still be used to refer to the relative velocity between the emitter and the reference frame in which the propagating medium is at rest. Equation (\ref{eqn:25}) is the counterpart of equation (\ref{eqn:23}) when the source aberration is taken into account, since it is obtained as 
\begin{equation}
f'=f-\frac{1}{2\pi}\mathbf{k}'\cdot\mathbf{v}_{rel}\text{,}\label{eqn:26}
\end{equation}
being $\mathbf{k}'$ the wave vector in the observer's location, given by equation (\ref{eqn:17}), and $\mathbf{v}_{rel}$ the relative velocity between the emitter and observer. 
Unlike equation (\ref{eqn:23}), the wave vector $\mathbf{k}'$ is not aligned with the line joining emitter and observer, but with the actual propagation direction of the wavefronts. Consequently, equation (\ref{eqn:26}) naturally incorporates the aberration produced by the source motion. The application of equation (\ref{eqn:26}) requires explicit determination of the local wave vector $\mathbf{k}'$, which is obtained precisely through the phase-function formalism developed in the previous sections.
In brief, equation (\ref{eqn:26}) then provides the most compact form of the classical DE, whereas equation (\ref{eqn:25}) corresponds to its explicit expression in terms of the geometrical variables of fig.\ref{Fig:09}. 

Thus, equation (\ref{eqn:26}) contains all the configurations discussed in the previous sections as a particular case. For example, for $\beta'_x = \beta'_\rho = 0$, no Doppler shift is obtained, whereas the standard expressions for moving emitters and moving observers are recovered under the corresponding limiting conditions.

\begin{figure*}[htbp]
\centering
\includegraphics[scale=1.2]{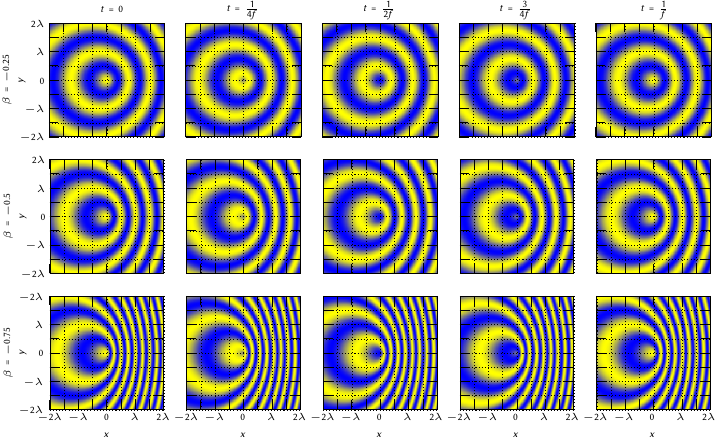}
\caption{Wavefronts resulting from drift associated with a moving medium in the emitter's reference frame (see supplementary videos 4 to 6). The sphericity of the wavefronts holds, but their centres shift with respect to the emitter's position. The negative values of $\beta$ mean that within this reference frame, the medium moves with velocity $-v\hat{\mathbf{x}}$.}
\label{Fig:10}
\end{figure*}

As an example, fig.\ref{Fig:10} shows the case where the propagating medium moves with respect to the emitter. The medium velocity is defined as $-v\hat{\mathbf{x}}$, and three different velocities are chosen, corresponding to $\beta=-0.25,-0.5,\text{and} -0.75$. These are indeed the cases shown in fig.\ref{Fig:04} but seen from the emitter's reference frame. Due to the moving medium, a drift of the wave fronts occurs. This drift promotes the gathering of wave fronts to the right of the emitter and its spreading to the left, or equivalently, different propagating velocities of the wave depending on direction. As has been mentioned above, despite these different propagating velocities, a stationary observer with respect to the emitter will not notice any frequency shift, since the wavelengths in the direction of the observation will enlarge or shorten accordingly.

\section{Conclusions}
Although classical DE has been known for more than a century, the present study shows that an explicit phase-based treatment not only reproduces the standard results but also provides a natural extension to more general configurations. In this regard,
the classical DE has been revisited through two complementary approaches: a geometrical construction and an explicit determination of the phase function of the propagating wave. For the case of a moving emitter and a stationary observer, both approaches lead to identical expressions for the frequency shift and aberration, providing a mutual verification of the obtained results.

The main contribution of this work is the explicit determination of the phase function associated with a moving source. Once this function is known, the local frequency and wave vector emerge naturally from its temporal and spatial derivatives. In this framework, the Doppler frequency shift is given by the temporal derivative of the phase, whereas aberration is determined by its spatial gradient. Consequently, both phenomena appear as different manifestations of the same underlying wave description.

The analysis also emphasizes the importance of the nonlinear dependence of the phase on time at the observer location. This dependence becomes particularly relevant in non-collinear configurations, where the observed frequency changes continuously as the relative geometry between source and observer evolves. The resulting expressions reduce to the widely known classical formulae in the collinear limit while providing a complete description for arbitrary observation directions.

The phase function yields a unified treatment of moving emitters, moving observers, and moving propagation media. In contrast with purely geometric methods, whose complexity rapidly increases in such circumstances, the phase-based approach yields a unified formulation in terms of relative positions and velocities. This formulation naturally incorporates the effects associated with a moving medium and demonstrates that the resulting frequency shift can be expressed in a form applicable to any reference frame.

In summary, the explicit construction of the phase function offers a unified and physically transparent description of the classical Doppler effect, linking frequency shift, aberration, and wave propagation within a common mathematical framework. Beyond its pedagogical value, this approach provides a useful tool for the analysis of more general propagation scenarios in which the source, observer, and medium may all be in relative motion. The phase function acts as a unifying element in the description of wave-propagation phenomena.

\section{Acknowledgments}
The authors would like to acknowledge the support of GIR Materiales Magn\'{e}ticos of the University of Valladolid and the fruitful discussions with its components that originated and shaped the contents of this work.

\appendix
\section{}\label{app:A}
By combining equations (\ref{eqn:03}) and (\ref{eqn:04}), it follows
\begin{equation}
\left(\frac{\sin\vartheta}{\sin\theta}\right)^2+\beta^2+2\beta\frac{\sin\vartheta}{\sin\theta}\cos\theta=1\text{,}
\end{equation}
which can be rewritten as a second degree equation in $\sin\vartheta$ as
\begin{equation}
\sin^2\vartheta+2\beta\sin\theta\cos\theta\sin\vartheta-\left(1-\beta^2\right)\sin^2\theta=0\text{,}
\end{equation}
one of whose solutions is the relationship shown in (\ref{eqn:05a}), that is,
\begin{equation}
\sin\vartheta=\sin\theta\left(\sqrt{1-\beta^2\sin^2\theta}-\beta\cos\theta\right)\text{,}
\end{equation}
This expression can be used to obtain that corresponding to $\cos\vartheta$. Indeed,
\begin{equation}
\begin{aligned}
\cos^2\vartheta&=1-\sin^2\vartheta=&\\
&=\beta^2\sin^4\theta+2\beta\cos\theta\sin^2\theta\sqrt{1-\beta^2\sin^2\theta}+\\
&\qquad+\cos^2\theta\left(1-\beta^2\sin^2\theta\right)=\\
&=\left(\beta\sin^2\theta+\cos\theta\sqrt{1-\beta^2\sin^2\theta}\right)^2\text{,}
\end{aligned}
\end{equation}
and finally,
\begin{equation}
\cos\vartheta=\beta\sin^2\theta+\cos\theta\sqrt{1-\beta^2\sin^2\theta}\text{.}
\end{equation}
as appeared in (\ref{eqn:05b})

\section{}\label{app:B}
The expression in (\ref{eqn:12}) can be rewritten as follows 
\begin{equation}
\begin{aligned}
1+&\frac{\beta}{1-\beta^2}\left(\beta+\frac{\cos\theta}{\sqrt{1-\beta^2\sin^2\theta}}\right)=\\
&=\frac{\sqrt{1-\beta^2\sin^2\theta}+\beta\cos\theta}{\left(1-\beta^2\right)\sqrt{1-\beta^2\sin^2\theta}}\text{.}
\end{aligned}
\end{equation}
After multiplying the numerator and the denominator by the factor $\sqrt{1-\beta^2\sin^2\theta}-\beta\cos\theta$, the result is as follows
\begin{equation}
\begin{aligned}
\frac{1-\beta^2\sin^2\theta-\beta^2\cos^2\theta}{\left(1-\beta^2\right)\sqrt{1-\beta^2\sin^2\theta}\left(\sqrt{1-\beta^2\sin^2\theta}-\beta\cos\theta\right)}=\\
=\frac{1}{1-\beta\left(\beta\sin^2\theta+\cos\theta\sqrt{1-\beta^2\sin^2\theta}\right)}\text{,}
\end{aligned}
\end{equation}
which is the factor appearing in (\ref{eqn:06}).

\section{}\label{app:C}
To check the relationship between equations (\ref{eqn:17}) and (\ref{eqn:18}), it is first necessary to determine the relationship between their moduli. Thus, 
\begin{equation}
\begin{aligned}
k'&=k\frac{\sqrt{\left(\beta\sqrt{1-\beta^2\sin^2\theta}+\cos\theta\right)^2+\left(1-\beta^2\right)^2\sin^2\theta}}{\left(1-\beta^2\right)\sqrt{(1-\beta^2\sin^2\theta}}=\\
&=k\frac{\sqrt{1-\beta^2\sin^2\theta+\beta^2\cos^2\theta+2\beta\cos\theta\sqrt{1-\beta^2\sin^2\theta}}}{\left(1-\beta^2\right)\sqrt{(1-\beta^2\sin^2\theta}}=\\
&=k\frac{\sqrt{1-\beta^2\sin^2\theta}+\beta\cos\theta}{\left(1-\beta^2\right)\sqrt{1-\beta^2\sin^2\theta}}\text{,}
\end{aligned}
\end{equation}
which, by virtue of the results in \ref{app:A} and \ref{app:B}, transforms into
\begin{equation}
k'=\frac{k}{1-\beta\left(\beta\sin^2\theta+\cos\theta\sqrt{1-\beta^2\sin^2\theta}\right)}=\frac{k}{1-\beta\cos\vartheta}\text{.}
\end{equation}
With regard to the direction of this vector, the tangent of the corresponding angle can be calculated as the quotient $\frac{k'_y}{k'_x}$, that is,
\begin{equation}
\frac{k'_y}{k'_x}=\frac{\left(1-\beta^2\right)\sin\theta}{\beta\sqrt{1-\beta^2\sin^2\theta}+\cos\theta}\text{.}
\end{equation}
By multiplying the numerator and the denominator by the factor $\sqrt{1-\beta^2\sin^2\theta}-\beta\cos\theta$ and simplifying the result, it follows that
\begin{equation}
\frac{k'_y}{k'_x}=\frac{\sin\theta\left(\sqrt{1-\beta^2\sin^2\theta}-\beta\cos\theta\right)}{\beta\sin^2\theta+\cos\theta\sqrt{1-\beta^2\sin^2\theta}}=\frac{\sin\vartheta}{\cos\vartheta}=\tan\vartheta\text{,}
\end{equation}
which shows that the wave vector $\mathbf{k}'$ is oriented along the direction given by the angle $\vartheta$.

\bibliographystyle{ieeetr}
\bibliography{main}

@book{Doppler:03,
  author={Christian Doppler},
  title={Ueber das farbige Licht der Doppelsterne und einiger anderer Gestirne des Himmels: Versuch einer das Bradley'sche Aberrations-Theorem als integrirenden Theil in sich schliessenden allgemeineren Theorie},
  publisher={K. B{\"o}hm Gesellschaft der Wissenschaften},
  year={1903 (original work published in 1842)}
}

@article{Bokor:09,
  title={A comparison of the electromagnetic and acoustic Doppler effects using geometrical diagrams},
  author={N\'{a}ndor Bokor},
  journal={Physics Education},
  volume={44},
  number={4},
  pages={368--373},
  year={2009},
  publisher={IOP Publishing Ltd}
}

@article{Michel:22,
  title={Galilean and relativistic Doppler/aberration effects deduced from spherical and ellipsoidal wavefronts respectively},
  author={Denis Michel},
  journal={Optik},
  volume={250},
  number={1},
  pages={168242},
  year={2022},
  publisher={Elsevier B.V. }
}

@article{Klinaku:21,
  title={The Doppler effect is the same for both optics and acoustics},
  author={Shukri Klinaku},
  journal={Optik},
  volume={244},
  pages={167565},
  year={2021},
  publisher={Elsevier B.V. }
}

@article{Klinaku:19,
  title={The Doppler effect and similar triangles},
  author={Shukri Klinaku},
  journal={Results in Physics},
  volume={12},
  pages={846--852},
  year={2019},
  publisher={Elsevier B.V. }
}

@book{halliday:13,
  title={Fundamentals of physics},
  author={David Halliday and Robert Resnick and Jearl Walker},
  year={2013},
  publisher={John Wiley \& Sons}
}

@book{serway:18,
  title={Physics for scientists and engineers},
  author={Raymond A. Serway and  John W. Jewett},
  year={2018},
  publisher={Cengage learning}
}

@book{tipler:22,
  title={Physics for scientists and engineers},
  author={Paul A. Tipler and Gene Mosca},
  year={2022},
  publisher={Macmillan Learning}
}

@book{cutnell:13,
  title={Introduction to Physics},
  author={John D. Cutnell and Kenneth W. Johnson},
  year={2013},
  publisher={John Wiley \& Sons, Inc.}
}

@book{pierce:19,
  title={Acoustics. An Introduction to Its Physical Principles and Applications},
  author={Allan D. Pierce},
  year={2019},
  publisher={Springer}
}

@book{whitham:74,
  title={Linear and Nonlinear Waves},
  author={Gerald Beresford Whitham},
  year={1974},
  publisher={Willey},
}

@book{morse:68,
  title={Theoretical Acoustics},
  author={Philip M. Morse and K. Uno Ingard},
  year={1968},
  publisher={Princeton University Press},
}

@book{rayleigh:96,
  title={The theory of Sound},
  author={John William {Strutt, Baron Rayleigh}},
  year={1896},
  volume={I--II},
  publisher={Mac Millan and Co.}
}

@book{blackstock:00,
  title={Fundamentals of Physical Acoustics},
  author={David T. Blackstock},
  year={2000},
  publisher={John Wiley \& Sons, Inc.},
}
\end{document}